\documentclass[conference]{IEEEtran}

\usepackage{cite}
\usepackage{url}
\usepackage[colorlinks=true,linkcolor=blue,citecolor=green]{hyperref}
\usepackage{graphicx}
\newcommand{\etal}{\textit{et al.}}
\usepackage{tabularx}
\usepackage{booktabs}
\usepackage{float}

\ifCLASSINFOpdf

\else

\fi

\begin{document}
%
\title{Do MPI Derived Datatypes Actually Help? A Single-Node Cross-Implementation Study on Shared-Memory Communication}

\author{\IEEEauthorblockN{Temitayo Adefemi}
\IEEEauthorblockA{School of Mathematics and Statistics\\
University of Edinburgh\\}}


%


\maketitle

\begin{abstract}
MPI's derived datatypes (DDTs) promise easier, copy-free communication of non-contiguous data, yet their practical performance remains debated and often reported only for a single MPI stack. This paper provides a cross-implementation assessment using three representative 2D applications—a Jacobi CFD solver, Conway's Game of Life, and a lattice-based image reconstruction—each written in two ways: (i) BASIC with manual packing/unpacking of non-contiguous regions and (ii) DDT using \texttt{MPI\_Type\_vector} / \texttt{MPI\_Type\_create\_subarray} with correct true extent via \texttt{MPI\_Type\_create\_resized}. For API parity, we benchmark identical communication semantics: non-blocking point-to-point (\texttt{Irecv}/\texttt{Isend}+\texttt{Waitall}), neighborhood collectives (\texttt{MPI\_Neighbor\_alltoallw}), and MPI-4 persistent operations (\texttt{*\_init}). We run strong and weak scaling on $1$--$4$ ranks, validating bitwise-identical halos, and evaluate on four widely used MPI implementations: \texttt{MPICH}, \texttt{Open MPI}, \texttt{Intel MPI}, and \texttt{MVAPICH 2} (single node, ARCHER2). 

\vspace{0.3cm}

\emph{Results are mixed.} DDTs can be fastest—e.g., non-blocking DDTs for the image reconstruction code on \texttt{Intel MPI}/\texttt{MPICH}—but can also be among the slowest on other stacks (\texttt{Open MPI}/\texttt{MVAPICH 2} for the same code). For the CFD solver, BASIC variants generally outperform DDTs across semantics, whereas for Game of Life, the ranking flips depending on the MPI library. We also observe stack-specific anomalies (e.g., \texttt{MPICH} slowdowns with DDT neighborhood/persistent modes). Overall, \emph{no strategy dominates across programs, semantics, and MPI stacks; performance portability for DDTs is not guaranteed}. The practical takeaway is to profile both DDT-based and manual-packing designs under the intended MPI implementation(s) and communication mode(s). Limitations include single-node scope and no memory-overhead analysis; multi-node studies and GPU-aware paths are promising directions for future work.
\end{abstract}

\vspace{0.3cm}


\section{Introduction}
The Message Passing Interface (MPI) has become a cornerstone of parallel programming since its first draft was presented at the Supercomputing '93 conference in November 1993~\cite{mpi1993}. 
Over time, the MPI standard has evolved to support a wide range of communication mechanisms between processes, including one-sided, synchronous, and asynchronous communication, as well as buffered communication~\cite{mpi30, mpi41}. 
Among its foundational features are \emph{derived datatypes}, also known as user-defined datatypes~\cite{gropp1999} \cite{mpi1994}. 

\vspace{0.3cm}

The derived datatype mechanism is a powerful and integral feature of MPI that enables communication of structured, non-contiguous, and non-homogeneous application data using any MPI communication operation. 
This eliminates the need for tedious, explicit, and often time and spaceconsuming manual packing and unpacking between intermediate communication buffers, thereby improving both efficiency and portability of MPI programs.

\vspace{0.3cm}

Although research has explored the performance efficiency of MPI-derived datatypes across various programs, most existing studies are confined to a single MPI implementation. 
As a result, the observations drawn from these experiments cannot be easily generalized to other implementations. 
To address this gap, our study aims to answer a critical research question: 
\emph{"Do MPI derived datatypes actually help across real codes and across multiple MPI stacks?"} 
We also seek to determine the extent of their benefits and limitations.

\vspace{0.3cm}

To this end, we implement two versions of three representative parallel programs to evaluate the use of MPI datatypes for memory and data management. 
The first version of each program employs standard MPI basic datatypes (e.g., \texttt{MPI\_INT}, \texttt{MPI\_DOUBLE}) along with manual packing and unpacking to manage communication buffers. 
The second version leverages advanced MPI-derived datatypes to represent non-contiguous memory regions directly, allowing communication without explicit data copying. 
For comprehensive evaluation and diversity in implementation, both versions of all programs are developed and tested using four widely adopted MPI implementations: 
\textbf{MPICH}, \textbf{Open MPI}, \textbf{Intel MPI}, and \textbf{MVAPICH 2}.

\vspace{0.3cm}

We evaluate two implementation strategies across three representative parallel programs: (i) a 2D single-phase CFD solver using the Jacobi iterative method, (ii) a 2D Conway's Game of Life cellular automaton, and (iii) a 2D lattice-based calculation for image reconstruction.

\vspace{0.3cm}

Each program is provided in two versions. The \textbf{BASIC} version uses only MPI basic types (e.g., \texttt{MPI\_INT}, \texttt{MPI\_DOUBLE}, \texttt{MPI\_CHAR}) and communicates non-contiguous data via explicit copies into contiguous scratch buffers. The \textbf{DDT} version replaces manual packing with MPI-derived datatypes (e.g., \texttt{MPI\_Type\_vector}, \texttt{MPI\_Type\_create\_subarray}, and \texttt{MPI\_Type\_create\_resized}) to describe halo rows, columns, and corners, enabling single-call exchanges.

\vspace{0.3cm}

Our results reveal no systematic performance advantage for either MPI-derived datatypes or manual packing. In some experimental configurations, derived datatypes yielded lower execution times, whereas in others they did not. This lack of a consistent trend holds across all communication semantics evaluated and all MPI implementations examined. The relative performance of derived datatypes versus manual packing and unpacking is highly context-dependent, influenced by factors such as problem size, the characteristics of the underlying computation, the chosen MPI stack, the communication semantics, and the overall communication volume in the application. Consequently, our only robust recommendation for practitioners seeking optimal runtime performance when faced with a choice between derived datatypes and manual packing/unpacking is to empirically benchmark and profile their specific application to make an informed decision.

\vspace{0.3cm}
The paper makes the following contributions:

\vspace{0.3cm}
\begin{itemize}
  \item We present a systematic, cross-implementation study of MPI-derived datatypes across three real stencil-style applications (a CFD solver, Conway's Game of Life, and an image reconstruction kernel), comparing them with equivalent manual packing using basic MPI types.

  \item We enforce strict API parity between the BASIC and DDT variants and evaluate three communication semantics---non-blocking point-to-point, neighborhood collectives, and MPI-4 persistent operations---to isolate the impact of the datatype representation itself from other algorithmic effects.

  \item We perform an extensive experimental evaluation across four widely used MPI implementations (MPICH, Open MPI, Intel MPI, and MVAPICH2) on a modern HPC system, considering both strong and weak scaling scenarios and validating bitwise correctness of the exchanged halo data.

  \item We analyze the resulting performance behavior and show that MPI-derived datatypes are neither universally beneficial nor universally harmful: their impact is strongly implementation- and semantic-dependent, highlighting the lack of performance portability and the necessity of empirical evaluation when choosing between BASIC and DDT designs.
\end{itemize}

\section{Related Work}
The MPI standard introduced the concept of \textit{derived datatypes} in its first version in the mid-1990s \cite{mpi1995}, recognizing them as a powerful feature for describing non-contiguous and heterogeneous data layouts. Derived datatypes allow an application to communicate complex memory layouts with a single MPI operation, instead of manually packing data into contiguous buffers \cite{carpen2017}. This mechanism improves programmer convenience and can enable optimized data transfer (e.g., using network scatter-gather capabilities) by the MPI library \cite{gropp1999, traff1999}. The elimination of tedious, error-prone manual packing and unpacking not only eases development but also offers potential performance benefits and improved code portability \cite{carpen2017}. Early sources, such as the MPI-1 standard and tutorials, emphasized the use of derived types for these reasons \cite{mpi1995}. Indeed, derived datatypes have been considered a cornerstone of MPI’s expressiveness since its inception.

\vspace{0.3cm}

Despite their promise, MPI-derived datatypes have historically faced performance challenges in practice. Many MPI implementations in the late 1990s and 2000s did not efficiently support derived types, leading to suboptimal performance. Gropp \etal{} noted as early as 1999 that users often found it faster to manually pack data into contiguous buffers than to rely on the MPI library’s handling of derived types \cite{gropp1999}. This ``manual packing'' workaround was widespread – it effectively defeated the purpose of derived datatypes, but was necessary because the library’s internal datatype processing could be slow \cite{gropp1999}. Early research, therefore, focused on improving MPI’s handling of derived datatypes. Träff \etal{} (1999) proposed a ``flattening on the fly'' technique to parse nested datatypes with an iterative, stack-based approach rather than expensive recursion \cite{traff1999}. Similarly, Gropp \etal{} (1999) provided a taxonomy of common non-contiguous access patterns and optimized internal representations for derived types, also using a stack-based parsing strategy \cite{gropp1999}. These optimizations reduced the overhead of interpreting complex datatypes inside MPI. Another seminal study by Reussner \etal{} (2000) extended the SKaMPI benchmarks to systematically test the performance of derived datatypes across different MPI libraries \cite{reussner2000}. Their results showed dramatic disparities between systems – for instance, on some platforms (Cray T3E) certain nested datatypes incurred up to a $16\times$ latency penalty compared to flat data, whereas another MPI (NEC SX-5) handled the same types much more efficiently (only $2–4\times$ overhead) \cite{reussner2000}. Such findings underscored that performance was not consistent across MPI implementations, and reinforced the need for careful, implementation-specific consideration when using derived datatypes.

\vspace{0.3cm}

In the 2000s, researchers continued to pursue higher performance and portability for MPI datatypes. Thakur, Gropp, and others introduced memory-conscious packing algorithms to accelerate non-contiguous data movement \cite{thakur2003}. For example, Byna \etal{} optimized the packing/unpacking of strided data by exploiting processor cache hierarchies, thereby reducing memory-access costs during communication \cite{thakur2003}. Their enhanced MPICH implementation could send matrix sub-blocks (a common non-contiguous pattern) significantly faster than both the default MPI routine and even user-written manual packing code \cite{thakur2003}. This was a striking result: with the right optimizations, derived datatypes \emph{can} outperform hand-pack code, contradicting the then-common assumption \cite{thakur2003}. Around the same time, researchers began to emphasize ``performance portability.'' Gropp \etal{} (2011) formalized a set of self-consistent performance guidelines for MPI derived datatypes \cite{gropp2011}. The idea was to specify how an ideal MPI implementation should perform relative to simpler operations – for example, that sending a non-contiguous message with a derived datatype should be no slower than an equivalent sequence of contiguous sends, or than using \texttt{MPI\_Pack} for the same data. These guidelines, if met, would assure users that they need not tailor their code to each MPI library for performance \cite{gropp2011}. However, their experimental tests across several MPI implementations revealed that many libraries violated these performance expectations \cite{gropp2011}. In other words, certain MPI stacks performed surprisingly poorly on some derived datatype scenarios, leading to ``unpleasant surprises'' for users \cite{gropp2011}. This finding echoed the long-standing concern that performance can vary widely between MPI vendors and versions. To mitigate this, others have proposed adaptive or just-in-time approaches. Prabhu and Gropp (2015) developed DAME, a runtime-compiled engine for MPI datatypes that can JIT-compile efficient packing code for a given data layout \cite{prabhu2015}. By generating optimized serialization routines on the fly, this approach aimed to combine portability with near hand-tuned performance on each architecture. Such efforts indicate the community’s ongoing commitment to making derived datatype performance both fast \emph{and} predictable.

\vspace{0.3cm}

Over the past 5–10 years, a number of studies have specifically compared derived datatype performance across multiple MPI implementations, highlighting both progress and remaining issues. Carpen-Amarie \etal{} (2017) performed an in-depth evaluation of several modern MPI libraries (including MPICH, Open MPI, and others) against a set of ``natural expectations'' for datatype handling \cite{carpen2017}. Their findings were ``in many ways surprising and disappointing'' \cite{carpen2017}. In some cases, using a derived datatype was actually slower than performing the equivalent pack and unpack with explicit MPI calls \cite{carpen2017}. This confirms that as of recent years, there are scenarios where MPI’s built-in handling still lags behind a manual approach – a situation one would hope MPI implementations would avoid. They also found that a fundamental MPI guarantee was not universally met: the MPI standard states that sending a message as a single contiguous datatype should perform the same as sending the equivalent number of individual basic elements, but in practice certain implementations showed a gap \cite{carpen2017}. The authors traced these issues to suboptimal decisions in the MPI libraries’ internal algorithms, such as simplistic heuristics at datatype commit time that do not cover all cases \cite{carpen2017}. Their extensive benchmarks, covering multiple data layouts and operations, demonstrated that no single MPI implementation was best in all cases, and each had particular weaknesses \cite{carpen2017}. These results reinforce the importance of testing across MPI stacks: performance conclusions drawn from one MPI (e.g., Open MPI) may not generalize to another (e.g., Intel MPI or MVAPICH 2). Indeed, our work is motivated by this gap in the literature – most prior studies evaluated the benefits of derived datatypes for only one MPI library. In contrast, cross-implementation studies (e.g., the above) show that their efficacy can differ widely.

\vspace{0.3cm}

Crucially, despite past performance pitfalls, many production scientific codes and libraries have embraced MPI derived datatypes to simplify data management and communication. The benefit is clear: a well-designed derived type can replace dozens of lines of packing logic with a single MPI call. For example, the \textit{waLBerla} multiphysics framework for lattice Boltzmann methods defines custom MPI datatypes to represent its ghost-layer regions (the halos at subdomain boundaries) \cite{walberla}. Using these types, waLBerla can update halo cells via one collective exchange without copying data element-by-element. Similarly, the \textit{LUDWIG} lattice-Boltzmann application employs non-blocking point-to-point calls with derived types to handle 3D halo swaps between subdomains \cite{akhmetova2017}. This allows moving all required boundary slices in one operation per neighbor, rather than many small messages. A study of LUDWIG’s communication notes that MPI datatypes provide a convenient way to treat non-contiguous boundary data ``as if it was contiguous'' for communication \cite{akhmetova2017}. Another prominent example is the heFFTe library (Highly Efficient FFT for Exascale). In parallel 3D FFT algorithms, each process must redistribute data (transpose) such that each gets the correct ``pencil'' sub-block for the next dimension’s transform. Traditional implementations like P3DFFT required packing these pencils into contiguous buffers before an all-to-all exchange. HeFFTe instead leverages MPI’s advanced datatype and collective features: it creates a 3D subarray datatype (via \texttt{MPI\_Type\_create\_subarray}) to describe the pencil layout, and then calls a single \texttt{MPI\_Alltoallw} to perform the global exchange of pencils \cite{ayala2022}. In essence, each rank tells MPI ``send these strided sub-blocks to the other ranks,'' and the MPI library handles the indexing and movement. This datatype-based approach avoids extra copies and has been documented as a high-performance strategy in recent FFT communication analyses \cite{ayala2022}. The caveat, as those studies point out, is that performance still hinges on the MPI implementation’s quality: for instance, some MPIs’ \texttt{Alltoallw} were found to be less optimized (and in one case not GPU-aware) compared to their regular \texttt{Alltoall}, which can diminish the benefit of using subarray datatypes on GPU-accelerated systems \cite{ayala2022}. Nevertheless, when the MPI library is efficient, applications like heFFTe achieve excellent scaling using derived types for communication, validating the productivity and performance advantages of the mechanism.

\vspace{0.3cm}

Finally, it is worth noting the ongoing advancements aimed at improving the performance of derived datatypes across all MPI stacks. Modern research projects have introduced zero-copy and hardware-accelerated communication techniques that specifically target non-contiguous data handling. Hashmi \etal{} (2020) present FALCON-X, a ``Fast and Low-overhead'' framework for MPI datatype processing that avoids intermediate copies by directly orchestrating data movement from source to destination buffers (even on GPUs) \cite{hashmi2020}. Their approach showed substantial latency improvements for irregular data patterns by exploiting features of both InfiniBand and GPU memory hierarchies. More recently, Chandrasekaran \etal{} (2022) investigated layout-aware, adaptive strategies for datatype communication within MVAPICH 2, including offloading packing to smart NICs or using CPU cores based on message characteristics \cite{chandrasekaran2022}. In their experiments, an optimized MVAPICH 2 with these strategies significantly outperformed several well-known MPI implementations. For instance, the adaptive scheme achieved up to a $3\times$ throughput improvement over Intel MPI 2019 and about 30\% over Open MPI 4.1 in specific non-contiguous transfer benchmarks \cite{chandrasekaran2022}. These results demonstrate that there is still room to improve the processing of derived datatypes markedly, and that newer techniques can achieve performance on par across different MPI stacks.
In summary, the literature over the past decades illustrates a trajectory from initial promise, through periods of inconsistent support, to a renewed focus on delivering both high performance and portability for MPI-derived datatypes. These works collectively suggest that when multiple MPI implementations are well optimized to handle derived datatypes, real applications \emph{do} benefit across the board – a premise that our study will further examine by evaluating derived datatype usage in representative codes on MPICH, Open MPI, Intel MPI, and MVAPICH2. Our research question, ``Do MPI-derived datatypes actually help across real codes and across multiple MPI stacks?'', directly builds on these prior observations: we aim to empirically determine if the modern MPI landscape has matured to the point where derived datatypes consistently realize their performance potential in practice, or if gaps remain between different implementations and use cases.

\section{Methodology}
We evaluate the two implementation strategies across three representative parallel programs: (i) a 2D single-phase CFD solver using the Jacobi iterative method, (ii) a 2D Conway's Game of Life cellular automaton, and (iii) a 2D lattice-based calculation for image reconstruction.

\vspace{0.3cm}

Each program is provided in two versions. The \textbf{BASIC} version uses only MPI basic types (e.g., \texttt{MPI\_INT}, \texttt{MPI\_DOUBLE}, \texttt{MPI\_CHAR}) and communicates non-contiguous data via explicit copies into contiguous scratch buffers. The \textbf{DDT} version replaces manual packing with MPI derived datatypes (e.g., \texttt{MPI\_Type\_vector}, \texttt{MPI\_Type\_create\_subarray}, with correct true extent via \texttt{MPI\_Type\_create\_resized}) to describe halo rows, columns, and corners, enabling single-call exchanges.

\vspace{0.3cm}

To ensure API parity, both BASIC and DDT variants use identical communication semantics. We benchmark three modes: non-blocking point-to-point (\texttt{Irecv/Isend+Waitall}), neighborhood collectives (\texttt{MPI\_Neighbor\_alltoallw}), and MPI-4 persistent operations (\texttt{..\_init}). Derived datatypes are constructed and committed once per run; \texttt{MPI\_Type\_commit} costs are excluded from per-step timings.

\vspace{0.3cm}

All codes use a 2D Cartesian process grid (\texttt{MPI\_Cart\_create}). Each rank stores a local subdomain padded with one-cell halos. To study size effects consistently. We run both strong scaling (fixed $N$, increasing ranks) and weak scaling (fixed local $n_{\mathrm{local}}$, increasing ranks) on 1-, 2-, and 4 processes. We run each variation six times and calculate the average to isolate noise, which is very particular in high-performance computing clusters.

\vspace{0.3cm}

We evaluate the performance of derived datatypes versus non-derived datatypes using distinct scaling strategies for each benchmark program. For weak scaling (where workload per process remains constant as processes increase), we vary problem complexity: the Image Reconstruction program uses 300, 600, and 900 iterations on a fixed image; the Computational Fluid Dynamics program uses 10,000, 20,000, and 30,000 iterations; and Conway's Game of Life uses landscape sizes of 1024, 2048, and 4096 with a constant iteration count. For strong scaling (where the total workload remains fixed as the number of processes increases), we maintain consistent problem sizes: the Image Reconstruction program uses 1000 iterations, the Computational Fluid Dynamics program uses 10,000 iterations on a fixed problem size, and Conway's Game of Life uses a landscape size of 1024.
This combination of minimal, moderate, and significant problem sizes enables a comprehensive analysis of how derived datatypes perform relative to non-derived datatypes under varying conditions, potentially revealing performance characteristics that would otherwise remain hidden in a single configuration. Table \ref{tab:scaling_params} summarizes the scaling parameters for each program.

\begin{table*}[t!]
\centering
\caption{Experimental Parameters for Scaling Analysis}
\label{tab:scaling_params}
\begin{tabularx}{\textwidth}{l X X} 
\toprule
\textbf{Program} & \textbf{Weak Scaling Parameters} & \textbf{Strong Scaling Parameters} \\
\midrule
Image Reconstruction & 300, 600, and 900 iterations & 1000 iterations (on a fixed problem size).  \\
\addlinespace 
Computational Fluid Dynamics & 10,000, 20,000, and 30,000 iterations. & 10,000 iterations (on a fixed problem size). \\
\addlinespace
Conway's Game of Life & Landscape sizes: 1024, 2048, 4096. \newline (Iterations constant) & Landscape size: 1024. \newline (Iterations constant) \\
\bottomrule
\end{tabularx}
\end{table*}

\vspace{0.3cm}

Correctness is validated by asserting bitwise identity of the halo contents between BASIC and DDT for several warm-up steps before timing.

\vspace{0.3cm}

We compile and run all variants on four widely used MPI stacks—\textbf{MPICH}, \textbf{Open MPI}, \textbf{Intel MPI}, and \textbf{MVAPICH 2} in Singularity containers using identical compiler flags for each program. All experiments were run on a single node on the ARCHER2 supercomputer located at the University of Edinburgh. Table \ref{tab:archer2_cache} shows the Cache Hierachy for ARCHER 2.

\begin{table}[H]
\centering
\caption{Cache Hierarchy for ARCHER2 CPU Nodes}
\label{tab:archer2_cache}
\begin{tabular}{ll}
\hline
\textbf{Cache Level} & \textbf{Size} \\
\hline
L1 Data Cache & 32 KiB per core \\
L1 Instruction Cache & 32 KiB per core \\
L2 Cache & 512 KiB per core \\
L3 Cache & 16 MiB per 8-core CCX \\
\hline
\end{tabular}
\end{table}

\newcolumntype{Y}{>{\raggedright\arraybackslash}X}

\begin{table*}[h!]
\centering
\footnotesize
\begin{tabularx}{\linewidth}{@{}l Y Y Y@{}}
\toprule
\textbf{Implementation} & \textbf{Base image} & \textbf{Tag / Version} & \textbf{Notes} \\
\midrule
MPICH     & \url{docker://mfisherman/mpich}                   & \url{4.2.0}                           &  \\
Intel MPI & \url{docker://intel/oneapi}                       & \url{2025.3.0-0-devel-ubuntu24.04}    & oneAPI toolchain tag\textsuperscript{*} \\
MVAPICH2  & \url{docker://nimbix/base-ubuntu-nvidia-mvapich2} & \url{latest}                          & resolves at build time\textsuperscript{*} \\
Open MPI  & \url{docker://mfisherman/openmpi}                 & \url{4.1.7}                           &  \\
\bottomrule
\end{tabularx}
\caption{Container images and tags used to build the MPI environments.}
\label{tab:mpi-versions}
\vspace{0.25em}
\raggedright\footnotesize \textsuperscript{*}\,For precise reproducibility, record the container digest and the MPI library version reported by \texttt{mpirun --version} inside each image.
\end{table*}

\section{Analysis}
\begin{figure*}[h!]
\centering
\includegraphics[width=1\textwidth]{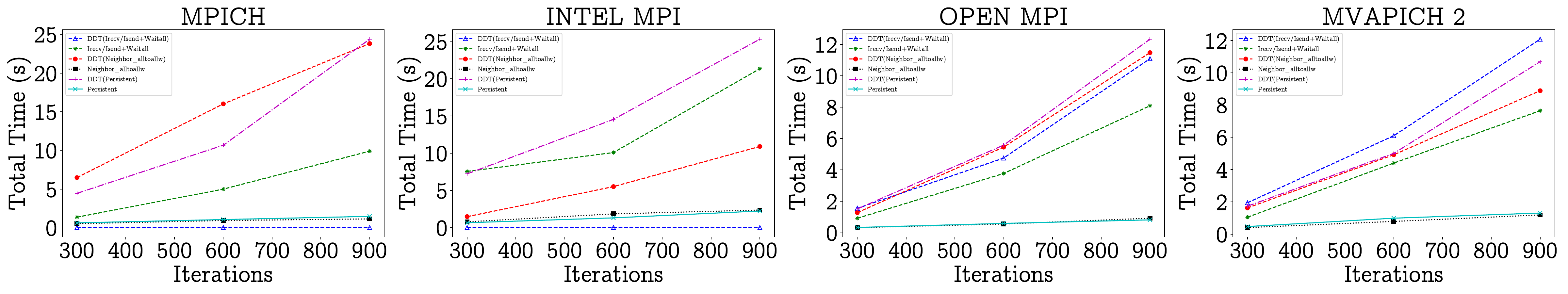}
\caption{Weak Scaling Experiments for the Image Reconstruction with Lattice Method Program}
\label{fig:ws_img}
\end{figure*}

\begin{figure*}[h!]
\centering
\includegraphics[width=1\textwidth]{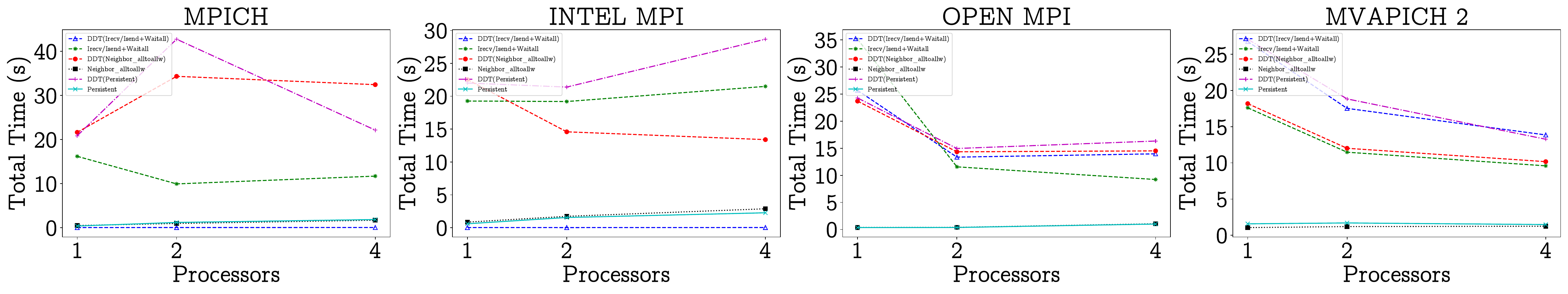}
\caption{Strong Scaling Experiments for the Image Reconstruction with Lattice Method Program}
\label{fig:ss_img}
\end{figure*}

\begin{figure*}[h!]
\centering
\includegraphics[width=1\textwidth]{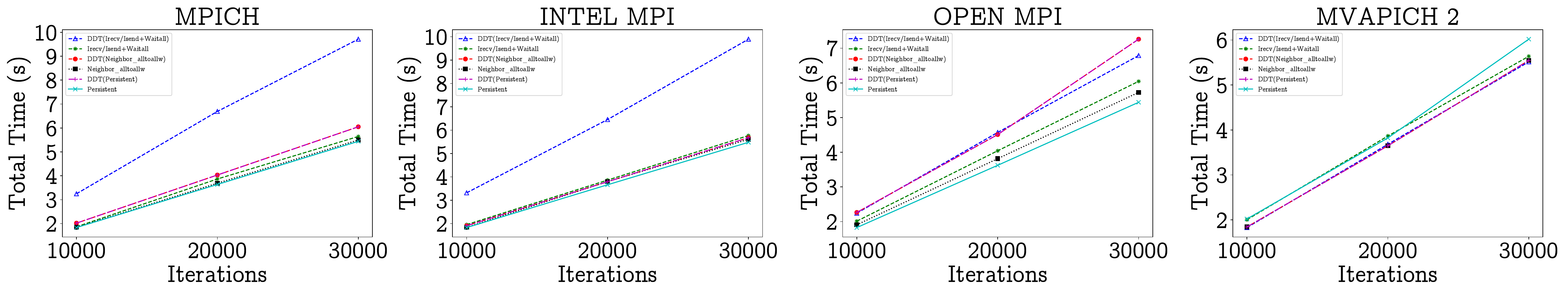}
\caption{Weak Scaling Experiments for Computational Fluid Dynamics Program}
\label{fig:ws_cfd}
\end{figure*}

\begin{figure*}[h!]
\centering
\includegraphics[width=1\textwidth]{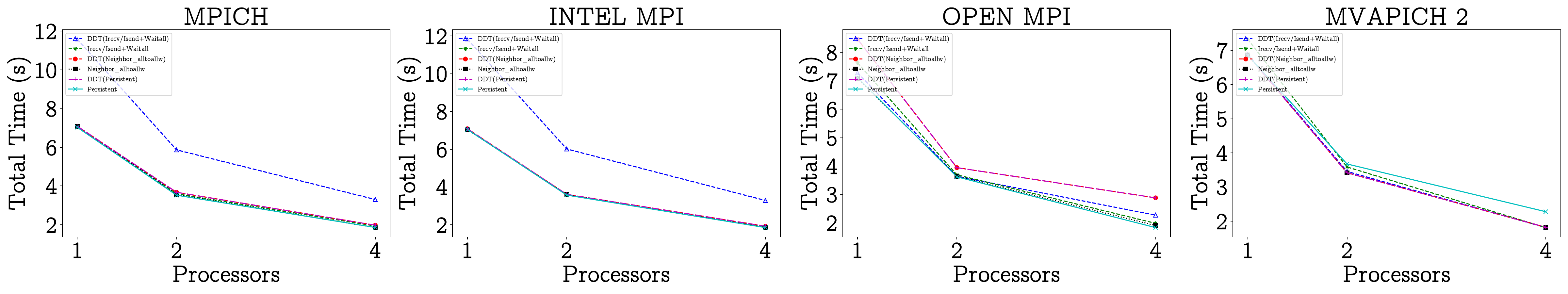}
\caption{Strong Scaling Experiments for the Computational Fluid Dynamics Program}
\label{fig:ss_cfd}
\end{figure*}

\begin{figure*}[h!]
\centering
\includegraphics[width=1\textwidth]{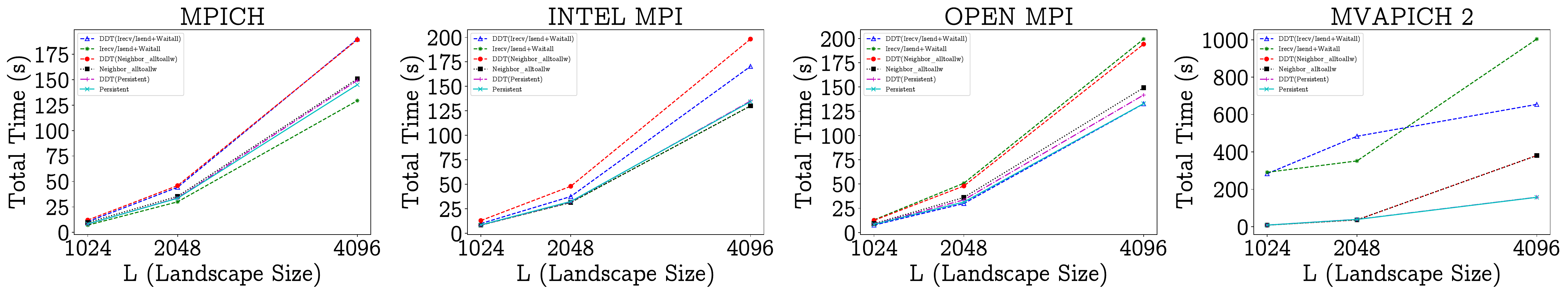}
\caption{Weak Scaling Experiments for Conway's Game of Life Program}
\label{fig:ws_gol}
\end{figure*}

\begin{figure*}[h!]
\centering
\includegraphics[width=1\textwidth]{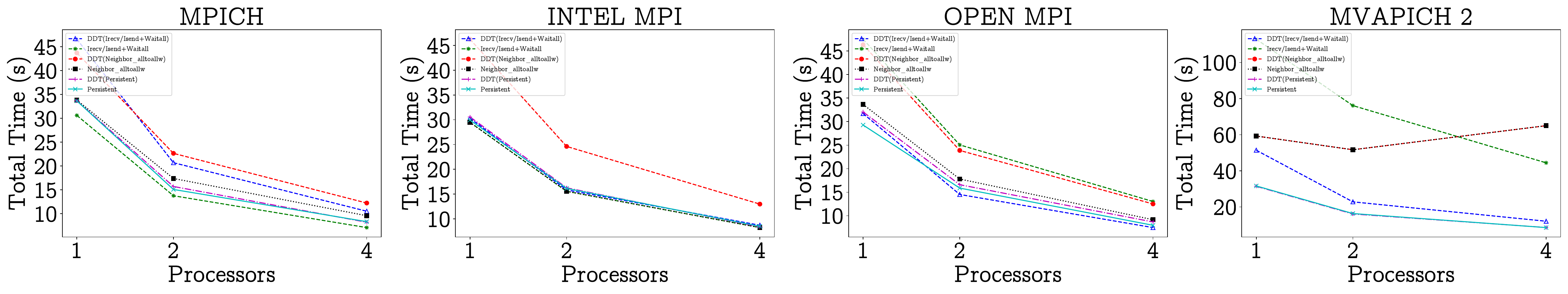}
\caption{Strong Scaling Experiments for Conway's Game of Life Program}
\label{fig:ss_gol}
\end{figure*}

The experiments produced mixed results; in some cases, using derived datatypes was more efficient, while in others, manual packing and unpacking was the better option. In this section, we will analyse each program separately, separating the weak and strong scaling analyses.

\subsection{Image reconstruction}
For the Image reconstruction program, which uses the lattice method, using derived datatypes with non-blocking point-to-point was the most efficient methodology across two MPI implementations (MPICH and Intel MPI). In strong scaling tests (\texttt{1000} iterations, \texttt{4} processors), this variant was high-speed, finishing in \texttt{0.0241} seconds on Intel MPI and \texttt{0.0417} seconds on MPICH. This is orders of magnitude faster than any other variant for those implementations. However, this same variant was less efficient than the other two MPI implementations, taking \texttt{13.97} seconds on Open MPI and \texttt{13.86} seconds on MVAPICH 2, making it one of the slowest variants in both.

\vspace{0.3cm}

Persistent communication with manual packing and unpacking  showed the most consistent performance across MPI implementations. In the \texttt{4}-processor strong scaling tests, the times were all in a relatively tight and efficient grouping: \texttt{1.01} seconds (Open MPI), \texttt{1.48} seconds (MVAPICH 2), \texttt{1.85} seconds (MPICH), and \texttt{2.27} seconds (Intel MPI).

\vspace{0.3cm}

Conversely, derived datatypes with neighborhood communication exhibited unexpected behavior in the strong-scaling experiments with MPICH. When increasing the processes from \texttt{1} to \texttt{2}, the program became more inefficient, slowing from \texttt{21.63} seconds on \texttt{1} process to \texttt{34.33} seconds on \texttt{2} processes (a \texttt{0.6301}x speedup). A similar behavior occurred with the implementation of derived datatypes with persistent communication (\texttt{..\_init}); this was more pronounced, slowing from \texttt{20.85} seconds on \texttt{1} process to \texttt{42.76} seconds on \texttt{2} processes (a \texttt{0.4876}x speedup). This behavior is odd and could be attributed to the problem's sparsity. When a problem size is granular enough to be handled by a single processor, the overhead of increasing the number of processes typically outweighs the benefits, as seen in this experiment. Intel MPI, Open MPI, and MVAPICH2 were all more efficient as the number of processes increased for some variants, such as the derived datatypes variant with non-blocking communication on Open MPI (\texttt{1} proc: \texttt{25.67}s, \texttt{2} procs: \texttt{13.36}s). For the Conway's Game of Life program and the Computational Fluid Dynamics program, all variants were more efficient as processes increased across all implementations, including MPICH, so the behavior of these two variants warrants further study.

\vspace{0.3cm}

The Image Reconstruction program's experiments continued to show atypical behavior. Not only did the derived datatypes (with neighborhood communication) and derived datatypes (with persistent communication) on MPICH become less efficient, but other variants showed little change. For example, the derived datatypes variant with non-blocking communication on Intel MPI showed almost no performance change, logging \texttt{0.0220} seconds for \texttt{1} process and \texttt{0.0241} seconds for \texttt{4} processes. Manual packing and unpacking variant with non-blocking communication was similar, moving from \texttt{19.27} seconds (\texttt{1} proc) to \texttt{21.50} seconds (\texttt{4} procs). It can be inferred that this behavior is due to the small number of iterations (\texttt{1000}) in the experiment's configuration. Other programs (Conway's Game of Life and Computational Fluid Dynamics) both use a sizable amount of iterations (\texttt{10,000}+), which allows performance improvement as processes increase.

\vspace{0.3cm}

To pinpoint a vital observation, when programs were executed on the Open MPI and MVAPICH 2 implementations, they often finished earlier than on the MPICH and Intel MPI implementations. For instance, in the manual packing and unpacking variant with non-blocking communication (\texttt{4} procs) test, Open MPI finished in \texttt{9.23} seconds and MVAPICH 2 in \texttt{9.59} seconds, while MPICH took \texttt{11.70} seconds. Intel MPI took \texttt{21.50} seconds. This conclusion, however, is inverted for the derived datatypes with a non-blocking communication variant, where Intel and MPICH were drastically faster.

\vspace{0.3cm}

Overall, the results are not conclusive; the derived data types variants did not consistently outperform the manual packing versions. For this program, there is no consistent pattern across the MPI implementations. For example, using derived datatypes with non-blocking communication on Intel MPI (\texttt{4} procs) is the fastest method at \texttt{0.024} seconds, but on Open MPI (\texttt{4} procs) it is one of the slowest at \texttt{13.97} seconds. This makes it impossible to draw meaningful conclusions about whether it is more efficient to use derived datatypes or to pack and unpack variables manually.

\subsection{Computational Fluid Dynamics}

The behavior of the Computational Fluid Dynamics program was more stable, with archetypal parallel-programming curves observed in both weak and strong scaling experiments across all implementations tested, suggesting that the large number of per-step iterations is responsible.

\vspace{0.3cm}

For example, in the weak scaling tests (fixed \texttt{4} processors), the time per iteration remained nearly constant. The derived datatypes variant that uses non-blocking communication clocked in at \texttt{0.1828} ms/iter for \texttt{10,000} iterations and \texttt{0.1839} ms/iter for \texttt{30,000} iterations. Similarly, the strong-scaling tests (fixed \texttt{10,000} iterations) showed excellent speedups. The manual and unpacking variant on the Intel MPI implementation dropped from \texttt{7.05} seconds on \texttt{1} processor to \texttt{1.89} seconds on \texttt{4} processors, achieving a \texttt{3.73}x speedup and \texttt{93.3}\% efficiency.
\vspace{0.3cm}

This experiment also provided more conclusive results. Manual packing and unpacking variants proved faster than derived datatypes for nearly all communication strategies evaluated.

\begin{itemize}
    \item Non-blocking (\texttt{Irecv/Isend+Waitall}): For Intel MPI on four processors, the manual and unpacking program was significantly faster (1.89 seconds) than the derived datatypes variant (3.29 seconds).

    \item Neighborhood (\texttt{MPI\_Neighbor\_alltoallw}): For Open MPI on four processors, the manual and unpacking program variant was much faster (1.91 seconds) than the derived datatypes variant (2.88 seconds).

    \item Persistent (\texttt{*\_init}): For MPICH on four processors, the manual and unpacking program(1.86 seconds) is faster than the derived datatypes variant (1.97 seconds).
\end{itemize}

\vspace{0.3cm} Derived datatypes with non-blocking point-to-point notably took significantly longer in the MPICH and Intel MPI implementations than other variants. For \texttt{10,000} iterations on \texttt{4} processors, this variant took \texttt{3.30} seconds for MPICH and \texttt{3.29} seconds for Intel MPI. These times are nearly twice those of their next-slowest variants (e.g., the implementation of derived datatypes using neighborhood communication on the MPICH implementation, clocked at \texttt{1.97} seconds). The performance degradation is possibly due to the constant back-and-forth of non-blocking communication, and the \texttt{Waitall} function can create overhead as the program has to wait for communication to complete before proceeding.

\vspace{0.3cm}

However, it is surprising that this deficiency is apparent only in the MPICH and Intel MPI implementations. The same variant shows highly competitive performance in the Open MPI and MVAPICH 2 implementations. For Open MPI, the time was \texttt{2.27} seconds (faster than its neighborhood communication variant), and for MVAPICH 2, it was one of the quickest variant at \texttt{1.81} seconds.

\vspace{0.3cm} 

Open MPI and MVAPICH2 were also more efficient than the MPICH and Intel MPI implementations in terms of peak performance. The best overall time for \texttt{10,000} iterations on \texttt{4} processors was from the manual packing and unpacking on the MVAPICH 2 implementation which clocked at \texttt{1.813} seconds, closely followed by the manual and unpacking variant, which uses persistent communication on Open MPI at \texttt{1.837} seconds. The best times for MPICH and Intel MPI were just behind at \texttt{1.849} seconds and \texttt{1.852} seconds, respectively.

\vspace{0.3cm} 

The experiment also reaffirmed that the mechanics and details of MPI implementations differ. Although they are all designed using the same Message Passing Interface, their performance behaviours are distinct, and it isn't reliable to infer your program's performance on another MPI implementation from data from a single implementation, especially as the program's scale increases. Although not tested in this experiment, the number of nodes would further distort the potential behavior in a different MPI implementation.

\subsection{Conway's Game of Life}
Conway's Game of Life also exhibited typical, predictable parallel-computing dynamics in the weak and strong scaling experiments. The program used a large landscape size and a fixed number of iterations (11,520) across all experiments.

\vspace{0.3cm}

In the MPICH and Intel MPI implementations, manual packing and unpacking with non-blocking communication were the most efficient. For a fixed landscape size of \texttt{1024} and \texttt{4} processors, this variant took \texttt{7.06} seconds for MPICH and \texttt{8.22} seconds for Intel MPI. Conversely, derived datatypes with neighborhood communication were the least efficient in these two implementations, taking \texttt{12.03} seconds with MPICH and \texttt{12.95} seconds with Intel MPI under the same conditions.

\vspace{0.3cm}

In contrast, for the Open MPI and MVAPICH 2 implementations, the manual packing and unpacking variant with non-blocking communication was the least efficient. For Open MPI, it was the slowest variant at \texttt{13.03} seconds with a problem size of (\texttt{1024}, \texttt{4} procs). For MVAPICH 2, this variant was a significant outlier, taking \texttt{291.05} seconds, while all other variants for a problem size of \texttt{1024} finished in under \texttt{13} seconds.

\vspace{0.3cm}

It is important to note that MPICH and Intel MPI performed similarly in several categories. For instance, in the manual packing and unpacking variant with persistent communication (L=\texttt{1024}, \texttt{4} procs), Intel MPI clocked in at \texttt{8.53} seconds and MPICH at \texttt{8.30} seconds. They were also close in the derived datatypes with persistent communication variant (Intel at \texttt{8.37}s, MPICH at \texttt{8.11}s). 

\vspace{0.3cm}

Based on the data, it is still very inconclusive whether using derived datatypes is more efficient than manually packing and unpacking variables. The results depend entirely on the specific MPI implementation. For example, with non-blocking communication (L=\texttt{1024}, \texttt{4} procs), Open MPI was fastest with derived datatypes (\texttt{7.49}s) and slowest without them (\texttt{13.03}s). The opposite was true for MPICH, which was fastest without derived datatypes (\texttt{7.06}s) and slower with them (\texttt{10.49}s).

\vspace{0.3cm}

The only universal conclusion we can draw from these results is that, across all MPI implementations and program variants, the program exhibited predictable parallel behavior. This implies that, as the landscape size increased while the number of processes remained fixed (weak scaling), the programs ran longer. For instance, the Intel MPI implementation with derived datatypes with persistent communication variant on \texttt{4} processors scaled from \texttt{9.88} seconds for L=\texttt{1024} to \texttt{170.37} seconds for L=\texttt{4096}.

\vspace{0.3cm}

Furthermore, as the number of processes used increased on a fixed landscape size (strong scaling), the programs ran faster. Using the same Intel MPI variant at L=\texttt{1024}, the total time dropped from \texttt{30.30} seconds on \texttt{1} processor to \texttt{15.83} seconds on \texttt{2} processors, and finally to \texttt{8.70} seconds on \texttt{4} processors, achieving a \texttt{3.48}x speedup. This behavior might be deemed expected and not worth discussing, but the Image Reconstruction program did not follow the same trend.

\section{Conclusion}
We investigated how derived datatypes affect program timing efficiency compared with manually packing and unpacking variables. To do this, we implemented six variants—using either derived datatypes or manual packing/unpacking—each combined with non-blocking, neighborhood, or persistent communication. These variants were evaluated across three parallel programs: (i) Conway’s Game of Life, (ii) an image reconstruction algorithm (lattice method), and (iii) a computational fluid dynamics application (Jacobi iterative method). The experiments were conducted on four MPI implementations: MPICH, Open MPI, MVAPICH2, and Intel MPI.

\vspace{0.3cm}

Our results show that the performance impact of derived datatypes versus manual packing/unpacking depends on several factors, including the nature of the program, problem size, communication semantics, communication volume, and the specific MPI implementation. As such, profiling remains essential for determining the most efficient choice for a given application.

\vspace{0.3cm}

This study has limitations. We did not examine the memory overhead associated with these approaches, and all experiments were conducted on a single node. Scaling to multiple nodes may introduce additional communication effects that could further influence performance. These aspects present opportunities for future research.


\newpage

\begin{thebibliography}{99}

\bibitem{mpi1995} MPI Forum, ``MPI: A Message-Passing Interface Standard, Version 1.1,'' 1995. Available at: \url{https://www.mpi-forum.org/docs/mpi-1.1/mpi-11-html/mpi-report.html}

\bibitem{mpi1993}
Message Passing Interface Forum, ``MPI: A Message Passing Interface,'' in \emph{Proceedings of Supercomputing '93}, Portland, Oregon, November 1993.

\bibitem{mpi1994}
Message Passing Interface Forum, ``MPI: A Message-Passing Interface Standard,'' Technical Report, University of Tennessee, Version 1.0, May 1994.

\bibitem{mpi30}
Message Passing Interface Forum, ``MPI: A Message-Passing Interface Standard Version 3.0,'' Technical Report, September 2012. Available at: \url{https://www.mpi-forum.org/}

\bibitem{gropp1999}
W. Gropp, E. Lusk, and A. Skjellum, \emph{Using MPI: Portable Parallel Programming with the Message-Passing Interface}, 2nd ed., MIT Press, 1999.

\bibitem{mpi41}
Message Passing Interface Forum, ``MPI: A Message-Passing Interface Standard Version 4.1,'' June 2023. Available at: \url{https://www.mpi-forum.org/}

\bibitem{traff1999} J. L. Träff \etal{}, ``Flattening on the Fly: Efficient Handling of MPI Derived Datatypes,'' EuroPVM/MPI 1999.

\bibitem{gropp1999} W. Gropp \etal{}, ``Improving the Performance of MPI Derived Datatypes,'' MPI Developer’s Conference, 1999.

\bibitem{thakur2003} R. Thakur, W. Gropp, and X. Sun, ``Improving the Performance of MPI Derived Datatypes by Optimizing Memory-Access Cost,'' IEEE/ACM Supercomputing, 2003.

\bibitem{gropp2011} W. Gropp \etal{}, ``Performance Expectations and Guidelines for MPI Derived Datatypes,'' EuroMPI 2011.

\bibitem{carpen2017} A. Carpen-Amarie \etal{}, ``On the Expected and Observed Communication Performance with MPI Derived Datatypes,'' EuroMPI/ACM Trans. Parallel Comput., 2016/2017.

\bibitem{reussner2000} R. Reussner \etal{}, ``A Benchmark for MPI Derived Datatypes,'' EuroPVM/MPI 2000. Available at: \url{https://dl.acm.org/doi/10.5555/648137.746641}

\bibitem{prabhu2015} T. Prabhu and W. Gropp, ``DAME: A Runtime-Compiled Engine for Derived Datatypes,'' EuroMPI 2015. Available at: \url{https://dl.acm.org/doi/10.1145/2802658.2802659}

\bibitem{hashmi2020} J. M. Hashmi \etal{}, ``FALCON-X: Zero-copy MPI Derived Datatype Processing on Modern CPU and GPU Architectures,'' J. Parallel Distrib. Comput., 2020. Available at: \url{https://www.sciencedirect.com/science/article/abs/pii/S0743731520302872}

\bibitem{chandrasekaran2022} K. K. Suresh \etal{}, ``Layout-Aware Hardware-Assisted Designs for MPI Derived Datatypes,'' IEEE Int’l Conf. Cluster Computing, 2022. Available at: \url{https://dl.acm.org/doi/abs/10.1177/1094342018808359}

\bibitem{akhmetova2017} D. Akhmetova \etal{}, ``Interoperability of GASPI and MPI in Large-Scale Scientific Applications,'' ISC 2017 (case study of LUDWIG). Available at: \url{https://www.pdc.kth.se/}

\bibitem{ayala2022} A. Ayala \etal{}, ``Performance Analysis of Parallel FFT on Large Multi-GPU Systems,'' IPDPS Workshops 2022 (discussion of \texttt{MPI\_Alltoallw} and subarray usage). Available at: \url{https://www.netlib.org/}

\bibitem{walberla} \textit{waLBerla} Framework Documentation – Ghost layer MPI datatypes. Available at: \url{https://walberla.net/}

\bibitem{akhmetova2017} D. Akhmetova \etal{}, ``Interoperability of GASPI and MPI in Large-Scale Scientific Applications,'' ISC 2017 (case study of LUDWIG). Available at: \url{https://www.pdc.kth.se/}
\end{thebibliography}
\end{document}